\documentclass{pasj00}
\draft
\begin{document}
\SetRunningHead{M. Shimizu et al.}{Fluctuation Amplitude from Cluster
  Abundance}   
\Received{2005/11/22} 
\Accepted{2006/2/1}

\title{Systematic bias in the estimate of cluster mass and the
fluctuation amplitude from cluster abundance statistics}

\author{Mamoru \textsc{Shimizu},\altaffilmark{1}
  
  Tetsu \textsc{Kitayama},\altaffilmark{2}
  
  Shin \textsc{Sasaki},\altaffilmark{3}
  
  and Yasushi \textsc{Suto}\altaffilmark{1}}

\altaffiltext{1}{Department of Physics, School of Science, The
  University of Tokyo, Tokyo 113-0033}

\altaffiltext{2}{Department of Physics, Toho University,  Funabashi,
  Chiba 274-8510}

\altaffiltext{3}{Department of Physics, Tokyo Metropolitan University,
  Hachioji, Tokyo 192-0397}

\email{mshimizu@utap.phys.s.u-tokyo.ac.jp, kitayama@ph.sci.toho-u.ac.jp,\\
  sasaki@phys.metro-u.ac.jp, suto@phys.s.u-tokyo.ac.jp}


\maketitle

\begin{abstract}
  We revisit the estimate of the mass fluctuation amplitude,
  $\sigma_{8}$, from the observational X-ray cluster abundance. In
  particular, we examine the effect of the systematic difference
  between the cluster virial mass estimated from the X-ray
  spectroscopy, $M_\mathrm{vir,\ spec}$, and the true virial mass of
  the corresponding halo, $M_\mathrm{vir}$. \citet{mazzotta04}
  recently pointed out the possibility that $\alpha_\mathrm{M} =
  M_\mathrm{vir,\ spec}/M_\mathrm{vir}$ is systematically lower than
  unity. We perform the statistical analysis combining the latest
  X-ray cluster sample and the improved theoretical models and find
  that $\sigma_8 \sim 0.76 \pm 0.01 + 0.50 (1-\alpha_{\mathrm{M}})$
  for $0.5\le\alpha_{\mathrm{M}}\le 1$, where the quoted errors are
  statistical only. Thus if $\alpha_{\mathrm{M}} \sim 0.7$, the value
  of $\sigma_8$ from cluster abundance alone is now in better
  agreement with other cosmological data including the cosmic
  microwave background, the galaxy power spectrum and the weak lensing
  data. The current study also illustrates the importance of possible
  systematic effects in mapping real clusters to underlying dark halos
  which changes the interpretation of cluster abundance statistics.
\end{abstract}

%%%%%%%%%%%%%%%%%%%%%%%%%%%%%%%%%%%%%%%%%%%%%%%%%%%%%%%%%%%%%%%%%%%%%%
\section{Introduction}
%%%%%%%%%%%%%%%%%%%%%%%%%%%%%%%%%%%%%%%%%%%%%%%%%%%%%%%%%%%%%%%%%%%%%%

Recent progress in observational cosmology has made it possible to
determine precise values of cosmological parameters including the
matter density parameter $\Omega_{\mathrm{M}}$, the cosmological
constant $\Omega_{\Lambda}$, the dimensionless Hubble constant
$h\equiv H_{0}/(100 \;\mathrm{km}\;\mathrm{s}^{-1} \;
\mathrm{Mpc}^{-1})$, and the mass fluctuation amplitude at
$8\;h^{-1}\;\mathrm{Mpc}$, $\sigma_8$. For instance, \citet{Spergel03}
obtained $(\Omega_\mathrm{M}, \, \Omega_{\Lambda}, \, h, \,
\sigma_{8})$ $= (0.29\pm0.07, \, 0.71\pm0.07, \, 0.72\pm0.05, \,
0.9\pm0.1)$ from the first-year data of the Wilkinson Microwave
Anisotropy Probe (WMAP) under the assumption of a spatially flat
universe, $\Omega_\mathrm{M} + \Omega_{\Lambda} = 1$. These estimates
slightly vary when combined with other observational probes such as
the galaxy power spectrum, weak lensing data and the Hubble diagram of
Type Ia supernovae, but the value of $\sigma_{8}$, which is the main
focus of the present paper, is always larger than 0.8. For instance,
\citet{Tegmark2004} concluded that $\sigma_8=0.89 \pm 0.02$ and
$\Omega_\mathrm{M}h = 0.213 \pm 0.023$.

Cluster abundance has been known as yet another useful probe of the
value of $\sigma_{8}$ \citep{White1993}. In a decade ago, the
methodology seemed to have almost established a value of $\sigma_8=0.9
\sim 1.0$ for standard $\Lambda$CDM (Lambda-dominated Cold Dark
Matter) cosmology (e.g., \cite{vl96,eke96,KS96,KS97,KSS98}) when
combined with a simple \textit{self-similar} model mass--temperature
(M-T) relation of X-ray clusters ($M \propto T^{3/2}$; see Kaiser
1986). \citet{seljak02}, however, showed that the use of the
\textit{observed} M-T relation \citep{finogu}, rather than the simple
self-similar M-T relation, leads to a much lower value:
%%%%%%%%%%%%%%%%%%%%%%%%%%%%%%%%%%%%%%%%%%%%%%%%%%%%%%%%%%%%%%%%%%%%
\begin{eqnarray}
  \sigma_{8}=
  (0.77\pm 0.07) (\Omega_\mathrm{M}/0.3)^{-0.44} (\Gamma/0.2)^{0.08},
\end{eqnarray}
%%%%%%%%%%%%%%%%%%%%%%%%%%%%%%%%%%%%%%%%%%%%%%%%%%%%%%%%%%%%%%%%%%%%
where $\Gamma$ is the shape parameter of the CDM power spectrum.

His result was also confirmed later by \citet{mshimizu03}. They
constructed the M-T relation from a combined analysis of X-ray
luminosity-temperature relation and temperature function of clusters,
and found $\sigma_8=0.7 \sim 0.8$. The derived M-T relation is
marginally consistent with the observed ones \citep{finogu,allen01},
but is in clear conflict with the simple relation as $M \propto
T^{3/2}$. Therefore it seems now clear that cluster abundance combined
with the reliable M-T relation consistently points to a systematically
lower value of $\sigma_{8}$ than the other cosmological indications.

\begin{figure}[tbh]
  \centering \FigureFile(100mm,100mm){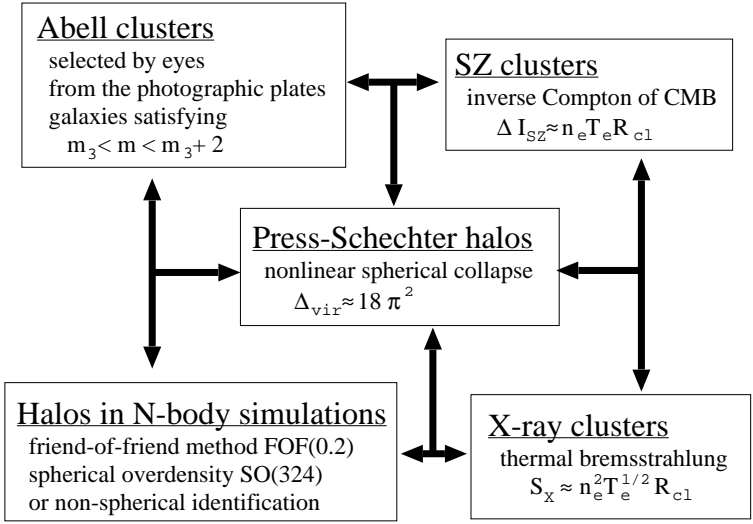}
  \caption{Relation among dark halos and galaxy clusters. 
    \label{fig:halocluster}}
\end{figure}

Given the above situation, it is important to recall that the
conventional modeling of galaxy clusters in terms of their mass or
temperature is oversimplified; non-sphericity, inhomogeneity and
substructure in the intracluster medium would give rise to both random
and \textit{systematic} variations to cluster properties with respect
to the simple model predictions. Figure~\ref{fig:halocluster} is the
improved version of the plot presented before by one of us
\citep{suto2002,suto2003}, which summarizes a wide range of practical,
and quite different, \textit{definitions} of dark halos that are
directly related to cosmology, but cannot be directly observed and of
galaxy clusters. Of course they are closely related, but any simple
one-to-one correspondence is unrealistic, and should be understood as
a working hypothesis. One has to improve the working hypothesis
continuously in order to increase the reliability of cluster abundance
statistics.

%%%%%%%%%%%%%%%%%%%%%%%%%%%%%%%%%%%%%%%%%%%%%%%%%%%%%%%%%%%%%%%%%%%%%%
%%%%%%%%%%%%%%%%%%%%%%%%%%%%%%%%%%%%%%%%%%%%%%%%%%%%%%%%%%%%%%%%%%%%%%

In this context, it is important to note that \citet{mazzotta04} and
\citet{rasia05} pointed out a potentially important source for the
systematic bias in estimating temperature and mass of X-ray clusters,
which motivated our current study. Indeed galaxy clusters may consist
of multi-phase temperature structure, and it is not a straightforward
task to define the \textit{overall} single temperature which
characterizes the cluster as a whole (see also \cite{vihklinin05}).

Usually X-ray observers estimate the temperature of a cluster by a
single temperature fit to the observed X-ray spectrum of the cluster.
Let us call it the \textit{spectroscopic} temperature,
$T_\mathrm{spec}$. It is natural, and indeed has been (implicitly)
assumed that $T_\mathrm{spec}$ is equivalent to the
\textit{emission-weighted} temperature, $T_\mathrm{ew}$, in which the
\textit{locally} defined temperature $T$ in a small region of the
cluster is averaged over with a weight of its emission measure. More
specifically, it is given by
%%%%%%%%%%%%%%%%%%%%%%%%%%%%%%%%%%%%%%%%%%%%%%%%%%%%%%%%%%%%%%%%%%%%%%%%%%%
\begin{eqnarray}
  \label{eq:T_ew}
  T_\mathrm{ew} \equiv \frac{\displaystyle\int T n^2 \Lambda(T) dV}
  {\displaystyle\int  n^2 \Lambda(T) dV} ,
\end{eqnarray}
%%%%%%%%%%%%%%%%%%%%%%%%%%%%%%%%%%%%%%%%%%%%%%%%%%%%%%%%%%%%%%%%%%%%%%%%%%%
where $n$ and $T$ are the gas density and temperature, $\Lambda(T)$ is
the cooling function ($\propto \sqrt{T}$ for thermal bremsstrahlung),
and the integration is over the entire cluster volume.
\citet{mazzotta04}, however, noticed that $T_\mathrm{spec}$
significantly underestimates the value of $T_\mathrm{ew}$ if clusters
have multi-phase temperature structure. Since relatively cool clumps
in a cluster exhibit many prominent emission lines, any single
temperature spectroscopic fitting to the cluster naturally tends to be
biased toward such low temperature clumps. In addition,
\citet{mazzotta04} showed that the systematic underestimate occurs
also in the case of thermal bremsstrahlung alone even without
considering contributions of emission lines. The authors introduced a
\textit{spectroscopic-like} temperature $T_\mathrm{sl}$:
%%%%%%%%%%%%%%%%%%%%%%%%%%%%%%%%%%%%%%%%%%%%%%%%%%%%%%%%%%%%%%%%%%%%%%%%%%%
\begin{eqnarray}
  \label{eq:T_sl}
  T_\mathrm{sl} \equiv \frac{\displaystyle\int T n^2 T^{-0.75} dV}
  {\displaystyle\int  n^2 T^{-0.75} dV} ,
\end{eqnarray}
%%%%%%%%%%%%%%%%%%%%%%%%%%%%%%%%%%%%%%%%%%%%%%%%%%%%%%%%%%%%%%%%%%%%%%%%%%%
which reproduces $T_\mathrm{spec}$ within a few percent for simulated
clusters hotter than a few keV (assuming Chandra or XMM-Newton
detector response functions). \citet{rasia05} performed a more
systematic study of the relation between $T_\mathrm{ew}$ and
$T_\mathrm{sl}$ using a sample of clusters from SPH simulations with
radiative cooling and heating, and found that $T_\mathrm{sl} =
(0.70\pm0.01) T_\mathrm{ew} + (0.29 \pm 0.05)~\mathrm{keV}$ for
$2~\mathrm{keV} \lesssim T_\mathrm{ew} \lesssim 13~\mathrm{keV}$. We
note here that \citet{mathiesen01}, based on their adiabatic
simulations, also noticed earlier that $T_\mathrm{spec}$ tends to be
lower than $T_\mathrm{ew}$, while the systematic difference is
somewhat smaller than that found by \citet{rasia05}.

As already discussed in \citet{rasia05}, the above result should have
a significant impact on the estimate of $\sigma_8$ from cluster
abundance. If one simply uses $T_\mathrm{sl}$, instead of
$T_\mathrm{ew} (> T_\mathrm{sl})$ in converting the temperature to the
underlying halo mass, one would underestimate the true mass and the
amplitude of the halo mass function, leading to an underestimation of
$\sigma_8$ as well. Furthermore, several numerical simulations
indicate that the assumption of hydrostatic equilibrium itself,
applied with the use of $T_\mathrm{ew}$, tends to underestimate the
cluster mass by $\sim 20\%$ (e.g., \cite{muanwong02, borgani04,
  rasia04}). Taken together, they may therefore account for the
systematically smaller value of $\sigma_8$ derived from cluster
abundance as described in the above. In reality, however, a reliable
prediction requires a more careful treatment of the selection function
and the statistical analysis of the observational sample. This is
exactly what we will conduct below.

We do not attempt to find the best-fit set of cosmological parameters,
but rather focus on the precise determination of $\sigma_8$. Thus in
this paper, we adopt a conventional $\Lambda$CDM model with the
following parameters; the matter density parameter
$\Omega_\mathrm{M}=0.27$, the cosmological constant
$\Omega_{\Lambda}=0.73$, and the dimensionless Hubble constant
$h_{70}=h/0.7=1$. We denote natural and decimal logarithms by $\ln$
and $\log$, respectively.

%%%%%%%%%%%%%%%%%%%%%%%%%%%%%%%%%%%%%%%%%%%%%%%%%%%%%%%%%%%%%%%%%%%%%%
\section{Method}
%%%%%%%%%%%%%%%%%%%%%%%%%%%%%%%%%%%%%%%%%%%%%%%%%%%%%%%%%%%%%%%%%%%%%%

The estimate of $\sigma_{8}$ from cluster abundance requires a variety
of theoretically and/or observationally calibrated relations among
mass, luminosity and temperature of clusters. Since our main interest
here is the effect of the difference between $T_\mathrm{ew}$ and
$T_\mathrm{spec}$, we carefully re-examine those relations that
directly involve the cluster temperature. Otherwise we adopt the
conventional modeling, following, but improving wherever possible, the
procedure of \citet{ikebe}.

\subsection{Mass function of dark matter halos \label{subsec:massfunc}}

Recent numerical simulations significantly advanced the understanding
of mass function of dark matter halos, and provide several fitting
formulae that are more accurate than their analytic counterpart
\citep{PS74}. In the present paper, we adopt the result of
\citet{jenkins}. The formula is based on the SO(324) halos which are
identified when their spherical over-density within the virial mass,
$M_\mathrm{vir}$, exceeds 324 times the mean matter background
density, $\bar\rho$:
%%%%%%%%%%%%%%%%%%%%%%%%%%%%%%%%%%%%%%%%%%%%%%%%%%%%%%%%%%%%%%%%%%%%%%%
\begin{eqnarray}
  \label{eq:mf_jenkins}
  \frac{dn}{d\ln M_{{\mathrm{vir}}}}
  &=&  0.316 \, \exp\left(-\left|\ln  \sigma^{-1}+0.67\right|^{3.82}\right)
  \frac{\bar{\rho}}{M_{\mathrm{vir}}}
  \frac{d\ln \sigma^{-1}}{d\ln M_{\mathrm{vir}}}, \\
  \sigma^{2}(M_{\mathrm{vir}}) &=& 4\pi\int P(k)
  \hat{W}^{2}(k; R_\mathrm{vir})  k^{2}dk ,\\
  \hat{W}(k; R_\mathrm{vir}) &=&\frac{3}{(kR_\mathrm{vir})^{3}}
  [\sin(kR_\mathrm{vir})-kR_\mathrm{vir}\cos(kR_\mathrm{vir})],
\end{eqnarray}
%%%%%%%%%%%%%%%%%%%%%%%%%%%%%%%%%%%%%%%%%%%%%%%%%%%%%%%%%%%%%%%%%%%%%%%
where $P(k)$ is the linear power spectrum of matter fluctuations, and
$R_\mathrm{vir} \equiv (3M_\mathrm{vir} /4\pi\bar{\rho})^{1/3}$.

\subsection{Mass-temperature relation of X-ray clusters}

The most uncertain procedure in estimating $\sigma_8$ from cluster
abundance is how to relate the dark matter halos, which are not
directly observable, to the actually observed X-ray clusters. Strictly
speaking, there is no reason why one can rely on any one-to-one
mapping between (simulated) halos and X-ray clusters; non-sphericity,
substructure, and merging history, among others, should be taken into
account to specify their individual properties (e.g., \cite{taruya00,
  komatsu01, suto2002, jingsuto02, suto2003, kitayama04}).
Nevertheless it is common to characterize halos and clusters merely as
a function of their mass and temperatures, respectively, and to relate
them on the basis of an empirically determined M-T relation (e.g.,
\cite{mshimizu03}). As described in Introduction, this procedure is
fairly successful. Nevertheless the resulting conclusion should be
interpreted with caution if one wants to take its precision and
accuracy seriously.

Let us make clear a few different definitions of mass and temperature
of clusters in order to specify our assumptions on their mutual
relation. For simulated clusters, one can compute the
emission-weighted temperature, $T_\mathrm{ew}$ (eq.~[\ref{eq:T_ew}]),
and the spectroscopic temperature, $T_\mathrm{spec}$. The mass of
observed clusters within a radius $R$ is usually derived on the
assumption of hydrostatic equilibrium:
%%%%%%%%%%%%%%%%%%%%%%%%%%%%%%%%%%%%%%%%%%%%%%%%%%%%%%%%%%%%%%%%%%%%%
\begin{equation}
  M(R) = - \frac{k_\mathrm{\scriptscriptstyle B} T_\mathrm{gas}(R) R}
  {G\mu m_\mathrm{\scriptscriptstyle H}} \left[ 
    \frac{d~\mathrm{ln}~n_\mathrm{gas}(R)}{d~\mathrm{ln}~R}
    + \frac{d~\mathrm{ln}~T_\mathrm{gas}(R)}{d~\mathrm{ln}~R}
  \right] ,
  \label{eq:xvir}
\end{equation}
%%%%%%%%%%%%%%%%%%%%%%%%%%%%%%%%%%%%%%%%%%%%%%%%%%%%%%%%%%%%%%%%%%%%%
where $k_\mathrm{\scriptscriptstyle B}$ is the Boltzmann constant, $G$
is the gravitational constant, $\mu$ is the mean molecular weight,
$m_\mathrm{\scriptscriptstyle H}$ is the proton mass, and
$n_\mathrm{gas}$ and $T_\mathrm{gas}$ are the gas density and
temperature, respectively. We define $M_\mathrm{vir,\ ew}$ and
$M_\mathrm{vir,\ spec}$ as those evaluated at a radius within which
the mean over-density is 324 when we use $T_\mathrm{ew}$ and
$T_\mathrm{spec}$, respectively, for $T_\mathrm{gas}$ in
equation~(\ref{eq:xvir}). To be more strict, our $T_\mathrm{ew}$ and
$T_\mathrm{spec}$ correspond to $T_\mathrm{ew}(R_\mathrm{vir})$ and
$T_\mathrm{spec}(R_\mathrm{vir})$ if the temperature profile is taken
into account.

For definiteness and simplicity, we assume that
%%%%%%%%%%%%%%%%%%%%%%%%%%%%%%%%%%%%%%%%%%%%%%%%%%%%%%%%%%%%%%%%
\begin{eqnarray}
  \label{eq:tspec-tew}
  T_\mathrm{spec} = \alpha_\mathrm{T} T_\mathrm{ew},
\end{eqnarray}
%%%%%%%%%%%%%%%%%%%%%%%%%%%%%%%%%%%%%%%%%%%%%%%%%%%%%%%%%%%%%%%%
where $\alpha_\mathrm{T}$ is a constant. According to \citet{rasia05},
$\alpha_\mathrm{T}\sim 0.7$ is favored from numerically simulated
clusters. Equations~(\ref{eq:xvir}) and (\ref{eq:tspec-tew}) alone
imply $M_\mathrm{vir,\ spec} = \alpha_\mathrm{T} M_\mathrm{vir,\ ew}$.
Taking account of the additional possibility that $M_\mathrm{vir,\
  ew}$ is systematically lower than the actual virial mass
$M_\mathrm{vir}$ of simulated clusters (e.g., \cite{muanwong02,
  borgani04, rasia04}), we simply relate $M_\mathrm{vir,\ spec}$ to
$M_\mathrm{vir}$ as
%%%%%%%%%%%%%%%%%%%%%%%%%%%%%%%%%%%%%%%%%%%%%%%%%%%%%%%%%%%%%%%%
\begin{eqnarray}
  \label{eq:mspec-mew}
  M_\mathrm{vir,\ spec} = \alpha_\mathrm{M} M_\mathrm{vir},  
\end{eqnarray}
%%%%%%%%%%%%%%%%%%%%%%%%%%%%%%%%%%%%%%%%%%%%%%%%%%%%%%%%%%%%%%%%
where the proportional constant $\alpha_\mathrm{M}$ accounts for both
the difference of $T_\mathrm{spec}$ and $T_\mathrm{ew}$ and that of
$M_\mathrm{vir,\ ew}$ and $M_\mathrm{vir}$. If $M_\mathrm{vir,\
  ew}=M_\mathrm{vir}$, then $\alpha_\mathrm{M} = \alpha_\mathrm{T}$.

Incidentally it is interesting to note that
equation~(\ref{eq:mspec-mew}) with $\alpha_\mathrm{M} \sim 0.6$
accounts for the well-known systematic difference between
$M_\mathrm{vir,\ spec}$ and the lensing mass estimate in galaxy
clusters (e.g., \cite{wu00, schmidt01}) if the latter is equal to the
actual virial mass. Another possible consequence of such a systematic
difference in temperature may be found in the interpretation of the
Sunyaev-Zel'dovich effect observations, where one needs to distinguish
the mass-weighted temperature and the X-ray emission-weighted or
spectroscopic temperature (e.g., \cite{yoshikawa00, yoshikawa01,
  komatsu2001}).

\begin{figure}[tbh]
  \FigureFile(80mm,80mm){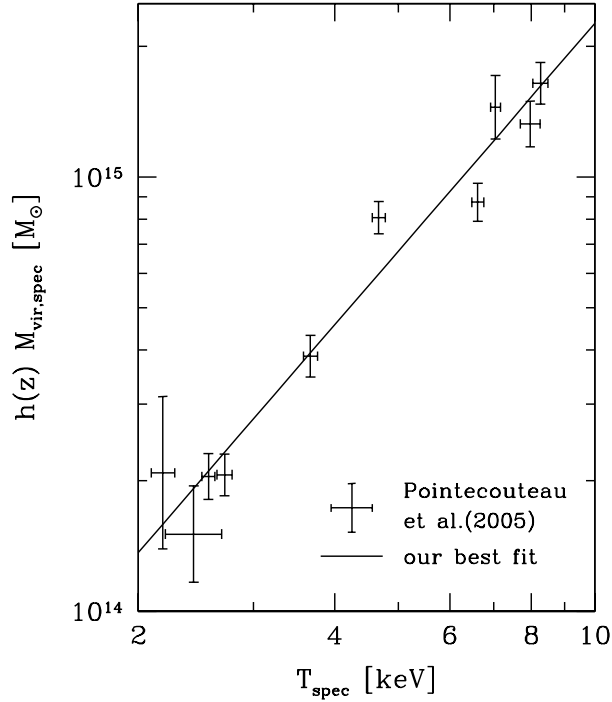} \centering
  \caption{Mass-temperature relation of the cluster sample by
    \citet{pointecouteau05} (symbols with error bars). Also plotted is
    the best-fit power-law that we adopt in the present analysis
    (eq.~[\ref{eq:mtfit}]). \label{fig:MT}}
\end{figure}

Next we fit the M-T relation to the cluster sample of
\citet{pointecouteau05} using a power-law model of the form:
%%%%%%%%%%%%%%%%%%%%%%%%%%%%%%%%%%%%%%%%%%%%%%%%%%%%%%%%%%%%%%%%
\begin{eqnarray}
  \label{eq:mtfit}
  \frac{M_\mathrm{vir,\ spec}}{10^{14}\; h(z)^{-1}\; M_{\odot}}
  =A_{\mathrm{vir}} 
  \left(
    \frac{T_{\mathrm{spec}}}{5\;\mathrm{keV}}
  \right)^{p},
\end{eqnarray}
%%%%%%%%%%%%%%%%%%%%%%%%%%%%%%%%%%%%%%%%%%%%%%%%%%%%%%%%%%%%%%%%
where
%%%%%%%%%%%%%%%%%%%%%%%%%%%%%%%%%%%%%%%%%%%%%%%%%%%%%%%%%%%%%%%%
\begin{eqnarray}
  h(z) = h_{70} \sqrt{\Omega_{\mathrm{M}}  (1+z)^{3}
    + (1-\Omega_{\mathrm{M}}-\Omega_{\Lambda}) (1+z)^{2}
    + \Omega_{\Lambda}
  } ~ .
\end{eqnarray}
%%%%%%%%%%%%%%%%%%%%%%%%%%%%%%%%%%%%%%%%%%%%%%%%%%%%%%%%%%%%%%%%
In practice, we follow \citet{mshimizu03}, and convert the values of
mass $\overline{M_{200}} \pm \Delta M_{200}$ in their paper to
$\overline{M_{\mathrm{vir}}} \pm \Delta M_{\mathrm{vir}}$ using the
redshift-dependent overdensity threshold estimated from the spherical
collapse model (e.g. \cite{KS96}), and assuming the universal density
profile of the hosting halo. We adopt that the inner power-law index
is equal to 1, and use the concentration parameter,
$\overline{c_{200}}$, listed in \citet{pointecouteau05}. In addition,
we scale the masses by $h(z)$, which corrects the
\textit{evolutionary} effect due to the observed redshifts of the
clusters following \citet{allen01} and \citet{arnaud05}. We perform
the fit on the $\log T$-$\log M$ plane, and find that
$A_{\mathrm{vir}} = 10^{0.829 \pm 0.018}$ and $p = 1.74 \pm 0.10$. The
best fit M-T relation and the observational data are plotted in
figure~\ref{fig:MT}.

%%%%%%%%%%%%%%%%%%%%%%%%%%%%%%%%%%%%%%%%%%%%%%%%%%%%%%%%%%%%%%%%%%%%%%
%%%%%%%%%%%%%%%%%%%%%%%%%%%%%%%%%%%%%%%%%%%%%%%%%%%%%%%%%%%%%%%%%%%%%%

\subsection{Fitting Procedure}

We search for the best-fit $\sigma_8$ from the maximum likelihood
analysis of X-ray cluster number density on the luminosity and
temperature plane, following \citet{ikebe}. We assume that at a given
spectroscopic temperature the cluster luminosity follows the log-normal
distribution:
%%%%%%%%%%%%%%%%%%%%%%%%%%%%%%%%%%%%%%%%%%%%%%%%%%%%%%%%%%%%%%%%%%%%%%%%%%%
\begin{eqnarray}
  \label{eq:pdf_L}
  p_L(\log L|T_\mathrm{spec}) d \log L
  =  \frac{1}{\sqrt{2\pi\sigma^{2}_{\log L}}}
  \exp
  \left\{
    -\frac{[\log L-
        \overline{\log {L}} (T_\mathrm{spec})]^{2}}
    {2\sigma_{\log L}^{2}}
  \right\}  d \log L ,
\end{eqnarray}
%%%%%%%%%%%%%%%%%%%%%%%%%%%%%%%%%%%%%%%%%%%%%%%%%%%%%%%%%%%%%%%%%%%%%%%%%%%
around the mean of the logarithm of the luminosity,
$\overline{\log{L}}(T_\mathrm{spec})$, for a given temperature
$T_\mathrm{spec}$, where $\sigma_{\log L}$ is its standard deviation.

The predicted number of clusters per unit logarithmic luminosity and
unit logarithmic temperature, $\mathcal{N}(L,T_{\mathrm{spec}})$, is
then given by
%%%%%%%%%%%%%%%%%%%%%%%%%%%%%%%%%%%%%%%%%%%%%%%%%%%%%%%%%%%%%%%%%%%%%%%%%%%
\begin{eqnarray}
  \label{eq:n_LT}
  &&  \mathcal{N}(L,T_{\mathrm{spec}}) \, d\log L \, d\log T_\mathrm{spec} \cr
  &=&
    \frac{dn}{dM_{\mathrm{vir}}}\Biggm|_{z=0.046} 
    \frac{dM_{\mathrm{vir}}}{dT_{\mathrm{spec}}}
    V_{\mathrm{max}}(L) p_L(\log L|T_{\mathrm{spec}}) 
    \frac{dT_{\mathrm{spec}}}{d\log T_{\mathrm{spec}}}  
    d\log L \, d\log T_\mathrm{spec}  ,
\end{eqnarray}
%%%%%%%%%%%%%%%%%%%%%%%%%%%%%%%%%%%%%%%%%%%%%%%%%%%%%%%%%%%%%%%%%%%%%%%%%%%
where we estimate the mass function at the median redshift value of
the cluster sample ($\langle z \rangle = 0.046$), and
$dM_{\mathrm{vir}}/dT_{\mathrm{spec}}$ is computed from
equation~(\ref{eq:mtfit}). The maximum comoving volume,
$V_{\mathrm{max}}(L)$, is given by
%%%%%%%%%%%%%%%%%%%%%%%%%%%%%%%%%%%%%%%%%%%%%%%%%%%%%%%%%%%%%%%%%%%%%%%%
\begin{eqnarray}
  \label{eq:vmax}
  V_{\mathrm{max}}(L) 
  = \frac{\Delta\Omega}{4\pi}\int^{\infty}_{0}
  dz \frac{dV}{dz} \int^{\infty}_{f_{\mathrm{lim}}}
  \frac{1}{\sqrt{2\pi\sigma_{\mathrm{f}}^{2}}}
  \exp\left[-\frac{(f-f_{0})^{2}}{2\sigma_{\mathrm{f}}^{2}}\right]df,
\end{eqnarray}
%%%%%%%%%%%%%%%%%%%%%%%%%%%%%%%%%%%%%%%%%%%%%%%%%%%%%%%%%%%%%%%%%%%%%
where $\Delta\Omega = 8.14\;\mathrm{sr}$ is the total sky coverage,
$dV/dz$ is a volume element per unit redshift per unit solid angle of
the sky, $f_{\mathrm{lim}} = 2.0\times 10^{-11} \; \mathrm{erg}\;
\mathrm{s}^{-1} \;\mathrm{cm}^{-2}$ is the observational flux limit in
the 0.1--2.4 keV band, $\sigma_{\mathrm{f}} = 10^{-12}\;
\mathrm{erg}\; \mathrm{s}^{-1}\; \mathrm{cm}^{-2}$ is a typical flux
measurement error. We set the average flux as $f_{0}=L/4\pi
d_{\mathrm{L}}^{2}(z)$, where $d_{\mathrm{L}}(z)$ is the luminosity
distance at $z$.

If the number of clusters with a given luminosity and temperature
obeys the Poisson distribution, the corresponding likelihood function
reduces to (e.g., \cite{cash})
%%%%%%%%%%%%%%%%%%%%%%%%%%%%%%%%%%%%%%%%%%%%%%%%%%%%%%%%%%%%%%%%%%
\begin{equation}
  \label{eq:likelihood}
  \ln\mathcal{L} = \sum_{i}\ln \mathcal{N}(L_{i},T_{i}) - \int
  \mathcal{N}(L,T_{\mathrm{spec}}) d\log L\; d\log T_{\mathrm{spec}} +
  \mbox{const.}, 
\end{equation}
%%%%%%%%%%%%%%%%%%%%%%%%%%%%%%%%%%%%%%%%%%%%%%%%%%%%%%%%%%%%%%%%%%
where $L_{i}$ and $T_{i}$ are the observed luminosity and
spectroscopic temperature of the $i$-th cluster, respectively. The
summation is taken over the observed clusters with temperature larger
than our adopted threshold $T_{\mathrm{min}}$. In practice, we
consider three values, $T_{\mathrm{min}}= 1.4$, $3.0$, and
$5.0\;\mathrm{keV}$ in order to see the extent to which our correction
for the flux-limit in the observational sample changes the conclusion
(see the next section). The integration in the second term of
equation~(\ref{eq:likelihood}) is performed for $41.5 < \log
(L/h^{-2}\mathrm{erg}\; \mathrm{s}^{-1}) <45.5$ and $ T_{\mathrm{min}}
< T_{\mathrm{spec}} < 11.2\; \mathrm{keV}$.

By maximizing the likelihood function given above, we are practically
fitting the observed X-ray temperature function (XTF) \textit{and} the
luminosity-temperature relation simultaneously. This method has the
following advantages over the conventional chi-square fitting to the
XTF data alone; 1) the Malmquist bias of the observed
luminosity--temperature relation is corrected using the observed XTF,
2) it is free from any bias arising from binning a small number of
data.

The above procedure is essentially the same as that of \citet{ikebe}
except that we adopt the observed M-T relation (eq.~[\ref{eq:mtfit}])
and the Jenkins mass function (eq.~[\ref{eq:mf_jenkins}]) in
equation~(\ref{eq:n_LT}). The integration over the temperature in
equation~(15) of \citet{ikebe} is omitted because the temperature
measurement errors are already incorporated in the observed M-T
relation. We have further checked that the fitted values of $\sigma_8$
will be unchanged within $\pm 0.01$ even if we artificially introduce
the intrinsic scatter of $\sqrt{\langle (\Delta T/T)^2 \rangle} \sim
0.05$ into the adopted M-T relation.

Since the main focus of the present paper is the systematic bias on
the value of $\sigma_8$ arising from errors in cluster mass
measurements, we do not intend to repeat time-consuming
multi-dimensional fit already performed by \citet{ikebe}. Instead, we
fix the luminosity--temperature relation of clusters. Specifically we
adopt the fitting result of \citet{ikebe}, PS(Flat,
$T>3\;\mathrm{keV}$) in their Table 3 where they assume the analytic
Press-Schechter mass function, the flatness of the universe and
$T_{\mathrm{spec}}>3\;\mathrm{keV}$:
%%%%%%%%%%%%%%%%%%%%%%%%%%%%%%%%%%%%%%%%%%%%%%%%%%%%%%%%%%%%%%%%%%%%%%%%
\begin{equation}
  \label{eq:LTobs}
  \overline{\log{L}_{\mathrm{0.1-2.4keV}}
    [{\mathrm{erg}\ \mathrm{s}^{-1}}]}
  (T_{\mathrm{spec}})
  = 42.19+2.44\log (T_{\mathrm{spec}}/1\mathrm{keV})-2\log h.
\end{equation}
%%%%%%%%%%%%%%%%%%%%%%%%%%%%%%%%%%%%%%%%%%%%%%%%%%%%%%%%%%%%%%%%%%%%%%%%
In the above expression,
$\overline{\log{L}_{\mathrm{0.1-2.4keV}}}(T_{\mathrm{spec}})$ is the
mean logarithmic X-ray luminosity of clusters in the 0.1--2.4~keV band
for $T_{\mathrm{spec}}$. We also adopt that $\sigma_{\log L}=0.23$
based on their results. We have made sure that the amplitude, the
slope and the scatter in the above fit are not affected by our use of
the Jenkins mass function and the observed M-T relation.

\begin{figure}[tbh]
  \FigureFile(80mm,80mm){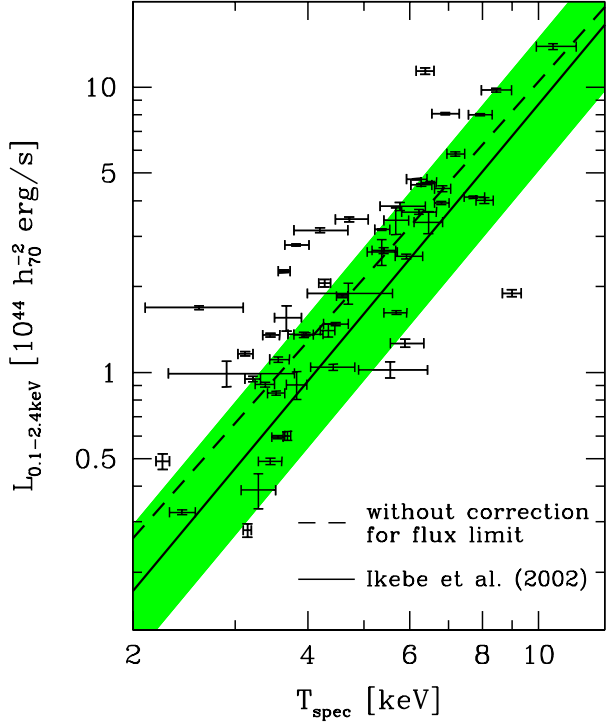} \centering \caption{The
    luminosity--temperature relations of clusters. Symbols with error
    bars represent the cluster sample of Ikebe et al. Solid line is
    our adopted power-law model (eq.~[\ref{eq:LTobs}]) with the
    corresponding log-normal errors (shaded region). Dashed line
    indicates the direct fit to the data without taking account of the
    Malmquist bias. \label{fig:LT}}
\end{figure}

Figure~\ref{fig:LT} shows the observed luminosity-temperature relation
of the sample \citep{ikebe}. The dashed line indicates the
straightforward power-law fit to those data. Our adopted
luminosity-temperature relation (solid line) takes into account the
flux-limit of the observations (i.e., the Malmquist bias), and thus
has a systematically lower amplitude than a direct fit to the observed
data.

%%%%%%%%%%%%%%%%%%%%%%%%%%%%%%%%%%%%%%%%%%%%%%%%%%%%%%%%%%%%%%%%%%%%%%
%%%%%%%%%%%%%%%%%%%%%%%%%%%%%%%%%%%%%%%%%%%%%%%%%%%%%%%%%%%%%%%%%%%%%%

\section{Results}

We perform the analysis for $\alpha_{\mathrm{M}} = M_\mathrm{vir,\
  spec}/M_\mathrm{vir} = 1$, 0.9, 0.8, 0.7, 0.6 and 0.5. The values of
$\sigma_{8}$ maximizing the likelihood function are summarized in
table~\ref{tbl:fitresults}. The clear general trend is that the
best-fit $\sigma_8$ systematically increases as $\alpha_{\mathrm{M}}$
decreases and $T_\mathrm{min}$ increases; if $\alpha_{\mathrm{M}}$
becomes smaller, the mass of the corresponding halo, $M_\mathrm{vir}$
is indeed larger than the naive estimate $M_\mathrm{vir,\ spec}$.
Therefore the resulting amplitude of the observed mass function
becomes larger, which requires larger $\sigma_8$. This is exactly
expected. On the other hand, the weak trend with respect to
$T_\mathrm{min}$, if real, might indicate that the sample is not yet
completely corrected for the Malmquist bias, and/or that the sample is
still statistically limited. Table~\ref{tbl:fitresults} indicates that
the best-fit value for $T_\mathrm{min}=3$~keV is roughly given as
%%%%%%%%%%%%%%%%%%%%%%%%%%%%%%%%%%%%%%%%%%%%%%%%%%%%%%%%%%%%%%%%%%%%%
\begin{eqnarray}
  \sigma_8 =  0.76 \pm 0.01 + 0.50(1-\alpha_{\mathrm{M}}).
\end{eqnarray}
%%%%%%%%%%%%%%%%%%%%%%%%%%%%%%%%%%%%%%%%%%%%%%%%%%%%%%%%%%%%%%%%%%%%%
The quoted errors represent the statistical error only, and the
systematic error due to the difference of $T_{\mathrm{min}}$ amounts
to $\pm 0.02$. Thus the systematic difference of the spectroscopic and
the true virial mass $\alpha_{\mathrm{M}} \equiv M_\mathrm{vir,\
  spec}/M_\mathrm{vir} \sim 0.7$ indeed reconciles the discrepancy of
$\sigma_{8}$ between the cluster abundance and Tegmark et al.'s
result, for instance.

%%%%%%%%%%%%%%%%%%%%%%%%%%%%%%%%%%%%%%%%%%%%%%%%%%%%%%%%%%%%%%%%%%%%%%% 
\begin{table}[thb]
  \caption{
    Best-fit values of $\sigma_{8}$
    for different 
    $\alpha_{\mathrm{M}}$ and $T_\mathrm{min}$.\label{tbl:fitresults}}  
   \begin{center}
     \begin{tabular}{cccc}\hline \hline
       $\alpha_{\mathrm{M}}$ & ${}>1.4\;\mathrm{keV}$
       & ${}>3\;\mathrm{keV}$ & ${}>5\;\mathrm{keV}$  \\\hline
       \# of clusters & 61 & 51 & 26 \\
       1   & $0.76^{+0.02}_{-0.01}$ & $0.78\pm0.02$          & $0.79\pm0.02$ \\
       0.9 & $0.79^{+0.02}_{-0.01}$ & $0.81^{+0.02}_{-0.01}$ & $0.82^{+0.02}_{-0.01}$\\
       0.8 & $0.83^{+0.02}_{-0.01}$ & $0.85\pm0.02$          & $0.86\pm0.02$\\
       0.7 & $0.88\pm0.02$          & $0.90\pm0.02$          & $0.91\pm0.02$\\
       0.6 & $0.94\pm0.02$          & $0.96\pm0.02$          & $0.97^{+0.02}_{-0.03}$\\
       0.5 & $1.02^{+0.03}_{-0.02}$ & $1.03^{+0.03}_{-0.02}$ & $1.04\pm0.03$
     \end{tabular}
   \end{center}
\end{table}
%%%%%%%%%%%%%%%%%%%%%%%%%%%%%%%%%%%%%%%%%%%%%%%%%%%%%%%%%%%%%%%%%%%%%% 

\begin{figure}[tbh]
  \centering \FigureFile(150mm,150mm){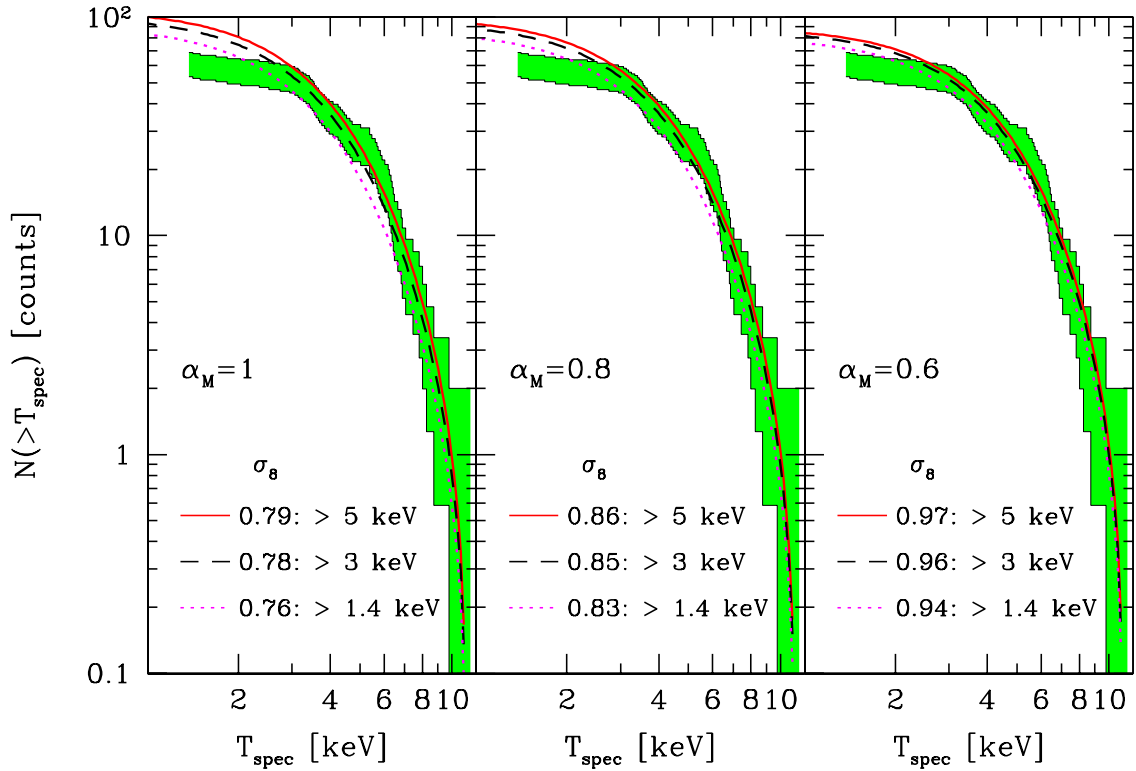}
  \caption{
    Cumulative number counts of the X-ray clusters as a
    function of $T_\mathrm{spec}$; \textit{Left}
    $\alpha_{\mathrm{M}}=1$, \textit{Middle} $\alpha_{\mathrm{M}}=0.8$
    \textit{Right} $\alpha_{\mathrm{M}}=0.6$.
    The shaded histogram indicates the range of the observational data
    of \citet{ikebe} with the Poisson errors. Solid, dashed and dotted
    curves show the results from our best-fit models for
    $T_\mathrm{min}= 5\; \mathrm{keV}$, 3 keV and 1.4 keV.
    \label{fig:bestfit}}
\end{figure}

To exhibit the goodness-of-fit of our derived parameters, we plot the
\textit{cumulative} number counts of clusters as a function of
$T_\mathrm{spec}$ in figure~\ref{fig:bestfit}:
\begin{equation}
  N(>T_{\mathrm{spec}}) =
  \int^{T=\infty}_{T= T_{\mathrm{spec}}}d\log T
  \int^{L=\infty}_{L=0} d\log L\
  \mathcal{N}(L,T).
\end{equation}
Given the simplified assumptions of single power-law fits both to the
observational M-T and to the underlying luminosity-temperature
relations, the fits are in reasonable agreement. Note that because the
horizontal axis of figure~\ref{fig:bestfit} is $T_\mathrm{spec}$, the
effect of $\alpha_{\mathrm{M}}$ does not look appreciable in the
resulting curves. In reality, however, the relation to the underlying
halo mass is very different, and this is why one needs larger values
of $\sigma_8$ to compensate the effect.

%%%%%%%%%%%%%%%%%%%%%%%%%%%%%%%%%%%%%%%%%%%%%%%%%%%%%%%%%%%%%%%%%%%%%%
%%%%%%%%%%%%%%%%%%%%%%%%%%%%%%%%%%%%%%%%%%%%%%%%%%%%%%%%%%%%%%%%%%%%%%

We would also like to call attention to the fact that the quality of
the fit degrades at $T_\mathrm{spec} < 3$ keV. This may possibly be
due to some unknown systematic effects in the observational sample of
\citet{ikebe}. While not obvious in the cumulative distribution like
figure~\ref{fig:bestfit}, there are few clusters observed around
$T_\mathrm{spec}\sim 5$ keV in their sample (see also Figure~6 of
\cite{mshimizu03}). Furthermore, the theoretical curves at
$T_\mathrm{spec} < 3$ keV are highly sensitive to how the flux limit
of the sample is corrected when modeling the XTF and the
luminosity--temperature relation. For these reasons, we prefer to rely
on the results of $T_\mathrm{min} > 3\; \mathrm{keV}$ even though the
difference due to the choice of $T_\mathrm{min}$ is not big
(table~\ref{tbl:fitresults}). Definitely, more reliable conclusions on
$\sigma_8$ should still need future well-controled statistical samples
of clusters.

%%%%%%%%%%%%%%%%%%%%%%%%%%%%%%%%%%%%%%%%%%%%%%%%%%%%%%%%%%%%%%%%%%%%%%
\section{Summary and discussion}
%%%%%%%%%%%%%%%%%%%%%%%%%%%%%%%%%%%%%%%%%%%%%%%%%%%%%%%%%%%%%%%%%%%%%%

We have shown that the systematic underestimate bias of the
spectroscopic and emission-weighted temperatures $\alpha_\mathrm{M}
\equiv M_\mathrm{vir,\ spec}/M_\mathrm{vir}$ may reconcile the
discrepancy of the values of $\sigma_{8}$ between the cluster
abundance and the other cosmological analyses if $\alpha_\mathrm{M}
\sim 0.7$.

Another equally important lesson that we have learned from this
analysis, however, is that the apparent agreement with independent
observational estimates does \textit{not} justify the use of any crude
but conventional assumptions. If we compare the value of $\sigma_8$
alone, the latest estimate $\sigma_8 \approx 0.9$ by WMAP is indeed in
good agreement with that obtained from cluster abundance argument a
decade ago \citep{vl96,eke96,KS96,KS97,KSS98}. Those previous analyses
relied on (i) a simple self-similar mass-temperature relation $M
\propto T^{3/2}$, (ii) a single phase of the intracluster temperature
(i.e., the spectroscopic temperature is identical to the
emission-weighted temperature of clusters), (iii) the analytical
Press--Schechter mass function of dark matter halos, and (iv) the
limited statistics of cluster abundance (e.g., the amplitude of
temperature function at 6 keV alone). The above assumptions have been
improved and updated for last several years, and now we know that the
previous assumptions (i) and (iii) systematically increased, while the
assumption (ii) decreased, the estimate of the value of $\sigma_8$, if
they are compared with their latest and improved counterparts.

It is often inevitable to introduce simple, reasonable, but
nevertheless inaccurate, assumptions in modeling and analyses of
cosmological observations since cosmological objects cannot be
predicted from the first principle of physics. Naturally we would like
to conclude that those assumptions are justified when they lead to the
same values of the cosmological parameters derived from independent
dataset and analysis. In most cases, the above procedure may not be
wrong, but still one has to keep in mind that the procedure as a whole
cannot be justified strictly because of the mere agreement since the
goal is \textit{not} to determine the precise value of parameters, but
to \textit{improve} the understanding of the underlying physical
processes involved.

In this sense, the estimate of $\sigma_8$ that we have presented in
this paper is certainly indicative, but may not be the final answer.
To reach more reliable conclusions, one needs to understand the
quantitative degree and the physical origin of the systematic
difference between spectroscopic and emission-weighted temperatures,
in addition to the improved statistical sample obviously. Furthermore,
the possible difference between the epochs of cluster formation and
observation may still be a source of systematic errors in the measured
$\sigma_{8}$ (\cite{KS97, ikebe}). Hopefully the recent progress of
numerical hydrodynamic simulations in cosmology and a variety of
proposals of galaxy surveys at intermediate and high redshifts will
significantly improve the situation in near future. For that
direction, the independent careful analysis of the systematics in the
Sunyaev-Zel'dovich and lensing cluster samples plays a very important
and complementary role.

\bigskip

We thank Eiichiro Komatsu and an anonymous referee for valuable
suggestions which improved the earlier manuscript of the present
paper. We are also grateful to Elena Rasia and Klaus Dolag for useful
comments in the early phase of this work, and to Gus Evrard and Joe
Mohr for discussions. This research was partly supported by
Grant-in-Aid for Scientific Research of Japan Society for Promotion of
Science (Nos.\ 14102004, 14740133, 15740157, 16340053).

%%%%%%%%%%%%%%%%%%%%%%%%%%%%%%%%%%%%%%%%%%%%%%%%%%%%%%%%%%%%%%%%%%%%%%%%%%%

%%%%%%%%%%%%%%%%%%%%%%%%%%%%%%%%%%%%%%%%%%%%%%%%%%%%%%%%%%%%%%%%%%%%%%%%%%%%%

\end{document}